\documentclass[12pt]{article}
\usepackage[dvips]{graphicx}
\usepackage{graphics}
\usepackage{amssymb}
\usepackage{amsmath,amssymb}
\usepackage{color} 
\usepackage{subfigure}
\usepackage[dvips]{graphicx}
\usepackage{graphics}
\usepackage{dcolumn}
\usepackage{amssymb}
\usepackage{bm}
\usepackage{amsmath}
\usepackage{color}

\def\spose#1{\hbox to 0pt{#1\hss}}

\def\lta{\mathrel{\spose{\lower 3pt\hbox{$\mathchar"218$}}
     \raise 2.0pt\hbox{$\mathchar"13C$}}}
\def\gta{\mathrel{\spose{\lower 3pt\hbox{$\mathchar"218$}}
     \raise 2.0pt\hbox{$\mathchar"13E$}}}
\newcommand{\be}{\begin{equation}}
\newcommand{\en}{\end{equation}}
\newcommand{\bea}{\begin{eqnarray}}
\newcommand{\ena}{\end{eqnarray}}

\begin{document}
\title{A boosted Kerr black hole solution
and the structure of a general astrophysical black hole}

\author{
Ivano Dami\~ao Soares \\
  Centro Brasileiro de Pesquisas F\'{\i}sicas, CBPF/MCTI \\Rio de Janeiro 22290-180, Brazil\\
  {\small email : ivano@cbpf.br}
  %
}

\date{}
%

\maketitle


\begin{abstract}
A solution of Einstein's vacuum field equations that describes
a boosted Kerr black hole relative to an asymptotic Lorentz frame at
the future null infinity is derived. The solution has three
parameters (mass, rotation and boost) and corresponds to the most
general configuration that an astrophysical black hole must have;
it reduces to the Kerr solution when the boost parameter is
zero. In this solution the ergosphere is north-south asymmetric, with
dominant lobes in the direction opposite to the boost. However
the event horizon, the Cauchy horizon and the ring
singularity {\bf --} which are the core of the black hole structure {\bf --} do
not alter, being independent of the boost parameter.
Possible consequences for astrophysical processes connected
with Penrose processes in the asymmetric ergosphere are discussed.

\end{abstract}


\maketitle

\section{Introduction and derivation of the solution}
\label{intro}
One of the most successful applications of General Relativity, the mathematical theory
of black holes, was developed based on the two exact black hole solutions described by the
Schwarzschild metric (obtained in 1915)\cite{schwarz} and the Kerr metric
(obtained in 1963)\cite{kerr0}. The Kerr metric describes a rotating black hole (with two
parameters, mass and angular momentum) and has the Schwarzschild black hole as its static limit configuration.
In particular the Kerr solution was of fundamental importance to the understanding of astrophysical processes
involved in objects with tremendous output of energy as quasars, pulsars and active galactic nuclei (AGNs).
Recent direct observations of the gravitational wave emission from a binary black hole
merger\cite{ligo}, with a mass ratio of the order of $0.8$, indicate that the resulting remnant black hole
is a Kerr black hole which must be boosted along a particular direction, with respect to the
asymptotic Lorentz frame at null infinity where such emissions have been detected. In this way
we have to add to the Kerr black hole description an additional
parameter -- the boost parameter -- connected to its motion with respect to the observation frame.
The boost of the remnant black hole is actually due to the presence of a nonzero net gravitational wave
momentum flux for the nonequal mass case of black hole collisions.
This must also be the case of astrophysical black holes in which a boost can be inherited
by the collapse of astrophysical objects with large bumps and other deformities.
As we will see the presence of this further parameter does not violate
theorems on the uniqueness of the structure of the Kerr black hole solution.
In the present paper we use geometric units $G=c=1$.
\par In the past literature several papers have dealt with the search of a boosted Kerr black
hole solution, the main objective of which was the use in $1+3$ Numerical Relativity
to obtain initial data for a system of spinning black holes. In \cite{karko} initial data for
a boosted Kerr black hole were constructed in the axially symmetric case with view to
a possible long-term numerical evolution of a single boosted black hole or a system of
black holes, and apparent horizons are found numerically. In \cite{scheel,matzner,matzner1}
the authors apply a Lorentz boost on the cartesian Kerr-Schild coordinates of a Kerr black hole
and carry out a $1+3$ decomposition, resulting in a slicing of the Kerr-Schild spacetime from which
black hole initial data can be propagated and apparent horizons are located.
In \cite{balasin,hogan} Lorentz boosts on Kerr-Schild coordinates of a Kerr geometry were
also used in evaluating a distributional energy-momentum with support in the singular region
of the metric as the basis for investigating ultra relativistic limit geometries.
These approaches are however distinct from the result in our paper
where we obtain an {\it exact stationary analytic solution} of a
boosted Kerr black hole relative to a Lorentz frame at future null infinity.
This exact stationary analytic solution should be expected to correspond to the final configuration
of the collision and merger  of a system of spinning black holes, therefore its importance
as a complementary test for the accuracy of the full numerical evolution of this system
up to the final remnant black hole.
\par The metric of a stationary boosted Kerr black hole is obtained here by using an integration
procedure analogous to that of Kerr in his original derivation of Kerr geometry\cite{kerr0}.
We make use of the simple and elegant apparatus presented in Sthephani et al.\cite{kramer}
(sections $29.1$ and $29.5$) for obtaining twisting Petrov D vacuum solutions
of Einstein's equations. We start with the metric expressed as
\begin{eqnarray}
\label{eqn1}
ds^2= 2 \omega^1 \omega^2 - 2 \omega^3 \omega^4
\end{eqnarray}
where the $1$-forms $\omega^a$ are given by
\begin{eqnarray}
\nonumber
\omega^1 &=&{\bar{ \omega}}^2 = - d \xi/{\bar{\rho}}P,\\
\omega^3 &=& du + L d\xi +{\bar{L}} d{\bar{\xi}},~~~~\omega^4 = dr + W d\xi +{\bar{W}} d{\bar{\xi}}+ H \omega^3,
\label{eqn2}
\end{eqnarray}
in Robinson-Trautman-type coordinates $(u,r,\xi,{\bar{\xi}})$\cite{rt}, where a bar
denotes complex conjugation. We also assume that
the metric functions are independent of the time coordinate $u$, namely, $\partial/\partial u$ is a Killing
vector of the geometry. $P$ is a real function. Einstein's vacuum equations lead then to~\cite{kramer}
\begin{eqnarray}
\rho^{-1}=-(r+i\Sigma),~~~~W=i~ \partial_{\xi} ~\Sigma \;,
\label{eqn3}
\end{eqnarray}
\begin{eqnarray}
H=\lambda/2- \frac{m r}{r^2+\Sigma^2}\;,
\label{eqn4}
\end{eqnarray}
\begin{eqnarray}
\lambda= 2 P^2 ~{\rm Re} ~(\partial_{\xi}\partial_{\bar \xi}~ \ln P)\;,
\label{eqn5}
\end{eqnarray}
\begin{eqnarray}
\lambda \Sigma+ P^2 {\rm Re}~ (\partial_{\xi}\partial_{\bar \xi}~\Sigma)=0\;,
\label{eqn6}
\end{eqnarray}
\begin{eqnarray}
2i \Sigma=P^2 (\partial_{\bar \xi}~L-\partial_{\xi} ~{\bar L})\;,
\label{eqn7}
\end{eqnarray}
where $m$ is a real constant and $\lambda=\pm 1$ is the curvature of the 2-dim surface
$d \xi d{\bar{\xi}}/P^2$. Here we adopt $\lambda=1$. The $r$-dependence is isolated in $\rho$ and $H$ so that the
remaining functions to be determined {\bf --} $P$, $\Sigma$ and $L$ {\bf --} are functions of
$(\xi,{\bar{\xi}})$ only.
We should mention that in equations (\ref{eqn3}) and (\ref{eqn4}) above we have further removed an integration
constant $r_0$ by a coordinate transformation that changes the origin of the affine
parameter $r$, setting thus effectively $r_0=0$. (cf. Section 29.1.4 of \cite{kramer}).
The integration of the field equations
reduces then to the integration of (\ref{eqn5}), (\ref{eqn6}) and (\ref{eqn7}).
For our purposes here we
will substitute the variables $(\xi,{\bar \xi})$ by $(\theta,\phi)$ via the stereographic transformation
\begin{eqnarray}
\nonumber
\xi=\cot (\theta/2)~ e^{i\phi}\;.
\label{eqn8}
\end{eqnarray}
\par From (\ref{eqn5}) we integrate the real function $P(\xi,{\bar \xi})$ by
assuming that $P$ has the form
\begin{eqnarray}
P=\frac{K(\theta,\phi)}{\sqrt{2}\sin^2(\theta/2)}\;.
\label{eqn9}
\end{eqnarray}
Eq. (\ref{eqn5}) results in
\begin{eqnarray}
1=K K_{\theta \theta}+K K_{\theta} \cot \theta-K^{2}_{\theta}+K^2+\frac{(K K_{\phi\phi}-K^{2}_{\phi})}{ \sin^2 \theta}\;.
\label{eqn10}
\end{eqnarray}
A general solution of (\ref{eqn10}) is given by
\begin{eqnarray}
K(\theta,\phi)= a + b~{\hat{\bf x}}\cdot {\bf n} \;, ~~~~~a^2-b^2=1\;,
\label{eqn11}
\end{eqnarray}
where ${\hat{\bf x}}=(\sin \theta \cos\phi, \sin \theta \sin \phi, \cos \theta)$ is
the unit vector along an arbitrary direction ${\bf x}$ and ${\bf n}=(n_1,n_2,n_3)$ is a constant unit vector
satisfying $n_1^2+n_2^2+n_3^2=1$. The solution (\ref{eqn11}) depends on three independent parameters
and defines a transformation of the generalized Bondi-Metzner-Sachs (BMS) group\cite{bondi1}
discussed by Sachs\cite{sachs1}, characterizing the general form of Lorentz boosts contained in the
homogeneous Lorentz transformations of the BMS group at null infinity.
In the original Kerr solution $K(\theta,\phi)=1$.
\par
Here our interest will be restricted to Lorentz boosts along the $z$-axis only,
\begin{eqnarray}
K(\theta)= a + b~ \cos \theta\;, ~~~~~a^2-b^2=1\;,
\label{eqn12}
\end{eqnarray}
that will correspond to a boosted axisymmetric solution. In the realm of black hole solutions,
(\ref{eqn12}) can be interpreted as corresponding to a Lorentz boost
of the black hole along the $z$-axis.
The boost parameter $\gamma$ parametrizes $a$ and $b$
as ($a=\cosh \gamma$,$~b=\sinh \gamma$), and is associated with the velocity $v= \tanh \gamma$
of the black hole relative to a Lorentz frame at future null infinity.
\par Assuming $\Sigma=\Sigma(\theta)$ and using (\ref{eqn12}),
eq. (\ref{eqn6}) in the variables $(\theta,\phi)$ reduces to
\begin{eqnarray}
\Sigma_{\theta \theta}+\cot \theta~ \Sigma_{\theta}+2 \frac{\Sigma}{K^2(\theta)}=0\;,
\label{eqn13}
\end{eqnarray}
yielding the regular solution
\begin{eqnarray}
\Sigma(\theta)= \omega ~\frac{(b + a~\cos \theta)}{(a+ b~\cos \theta)}\;,
\label{eqn14}
\end{eqnarray}
where $\omega$ is an arbitrary constant, to be identified with the rotation
parameter of the solution. Using (\ref{eqn14}) we now integrate
Eq. (\ref{eqn7}). We accordingly adopt
\begin{eqnarray}
L=i \mathcal{L}(\theta) e^{-i \phi}\;,
\label{eqn15}
\end{eqnarray}
resulting in
\begin{eqnarray}
{\mathcal{L}}_{\theta}- {\mathcal{L}}/\sin \theta + (1-\cos \theta)~\frac{\Sigma(\theta)}{K^{2}(\theta)}=0 \;.
\label{eqn16}
\end{eqnarray}
A general solution for (\ref{eqn16}) is given by
\begin{eqnarray}
\label{eqn17}
{\mathcal{L}}({\theta})=\Big( \frac{1-\cos\theta}{\sin \theta}\Big)\Big\{C_1  -\frac{\omega}{2 b^2}\Big(\frac{a^2+2 a b \cos \theta+b^2}{(a+b \cos \theta)^2} \Big)\Big \} \;,
\end{eqnarray}
where $C_1$ is an arbitrary constant. The apparent singular behaviour of the above solution for a zero boost ($b^2=0$) can be
eliminated either by fixing $C_1=\omega/2 b^2$ or by a coordinate transformation on the final metric.
In both cases it finally results
\begin{eqnarray}
\label{eqn18}
{\mathcal{L}}({\theta})=-\omega \frac{(1-\cos\theta)~ \sin \theta}{2 (a+ b\cos \theta)^{2}}= -2 \omega ~\frac{\cos \theta/2 ~\sin^{3}\theta/2}{(a+b \cos \theta)^2} \;.
\end{eqnarray}
\par Also the equation defining $W$ yields
\begin{eqnarray}
\label{eqn19}
W= i \omega ~\frac{\sin \theta}{(a+ b\cos \theta)^{2}}~(\sin^{2}{\theta/2})~ e^{-i\phi}  \;,
\end{eqnarray}
implying that
\begin{eqnarray}
\label{eqn20}
W d\xi+{\bar W}d{\bar \xi}= - ~\frac{\omega}{(a+ b\cos \theta)^{2}}~\sin^{2}\theta ~d\phi  \;.
\end{eqnarray}
Analogously
\begin{eqnarray}
\label{eqn21}
L= - i ~\frac{2\omega}{(a+ b\cos \theta)^{2}}~ \cos \theta/2~ \sin^3 \theta/2 ~e^{-i\phi}
\end{eqnarray}
results in
\begin{eqnarray}
\label{eqn22}
L d\xi+{\bar L}d{\bar \xi}=  ~\frac{\omega}{(a+ b\cos \theta)^{2}}~\sin^{2}\theta ~d\phi \;.
\end{eqnarray}
We also obtain
\begin{eqnarray}
\label{eqn22}
\frac{1}{\rho}=-(r+i\Sigma)=-  ~\Big(r + i \omega~\frac{b+a \cos \theta}{a+ b \cos \theta} \Big) \;.
\end{eqnarray}
The metric (\ref{eqn1}) results finally
\begin{eqnarray}
\label{eqn23}
\nonumber
ds^2&=&\frac{r^2+\Sigma^2(\theta)}{(a+b \cos \theta)^2} (d \theta^2+ \sin^2\theta d\phi^2)-\\
\nonumber
&-&2\Big(du+\frac{\omega \sin^2\theta}{(a+b \cos \theta)^2} d\phi \Big) \Big(dr-\frac{\omega \sin^2\theta}{(a+b \cos \theta)^2}d \phi\Big) \\
&-&\Big(du+\frac{\omega \sin^2\theta}{(a+b \cos \theta)^2} d\phi \Big)^2 \frac{1}{r^2+\Sigma^2(\theta)} \Big(r^2-2mr+ \Sigma^2(\theta) \Big) \;,
\end{eqnarray}
\\
where $\Sigma(\theta)= \omega ({b+a\cos \theta})/({a+b\cos \theta})$. This metric describes
a boosted Kerr black hole, the boost being along its axis of rotation, with respect to an asymptotic
Lorentz frame at future null infinity. For $b=0$ the metric (\ref{eqn23}) is the Kerr metric
in retarded Robinson-Trautman coordinates. For $\omega=0$ it represents a boosted
Schwarzschild black hole along the $z$-axis\cite{bondi1}.

\section{Properties of the solution: the ergosphere and horizons}
\label{sec:1}

A direct examination of (\ref{eqn23}) shows that $\partial/\partial u$ and $\partial/\partial \phi$ are
Killing vectors of the geometry. The boosted Kerr geometry also presents an ergosphere, defined by the limit surface
for static observers, namely, the locus where the Killing vector $\partial/\partial u$
becomes null\cite{poisson}. In the coordinate system of (\ref{eqn23}) the equation for
the static limit surface $g_{uu}=0$ results in
\begin{eqnarray}
\label{eqn24}
r^2-2 m r + \Sigma^2(\theta)=0 \;,
\end{eqnarray}
\\
namely,
\begin{eqnarray}
\label{eqn24}
r_{\rm stat}(\theta)=m+ \sqrt{m^2- \omega^2 \Big(\frac{b+a\cos \theta}{a+b\cos \theta}\Big)^2} ~\;.
\end{eqnarray}
\\
The horizons of the boosted Kerr metric are the surfaces where
\begin{eqnarray}
\label{eqn25}
g^{rr}=\frac{r^2 -2 m r + \omega^2~(\sin^2\theta+ (b+a \cos \theta)^2)/K^2(\theta)}{r^2+\Sigma^2(\theta)}=0 \; ,
\end{eqnarray}
resulting, after some algebra,
\begin{eqnarray}
\label{eqn26}
r_{\pm}= m \pm \sqrt{m^2-\omega^2} \; .
\end{eqnarray}
We see that the event horizon $r_{+}$ and the Cauchy horizon $r_{-}$ do not alter
by the effect of the boost. This should be expected since both horizons are lightlike surfaces
and therefore Lorentz invariant, contrary to the case of the static limit surface.
\par
\begin{figure}
\begin{center}
  \includegraphics[width=7.55cm,height=8.0cm]{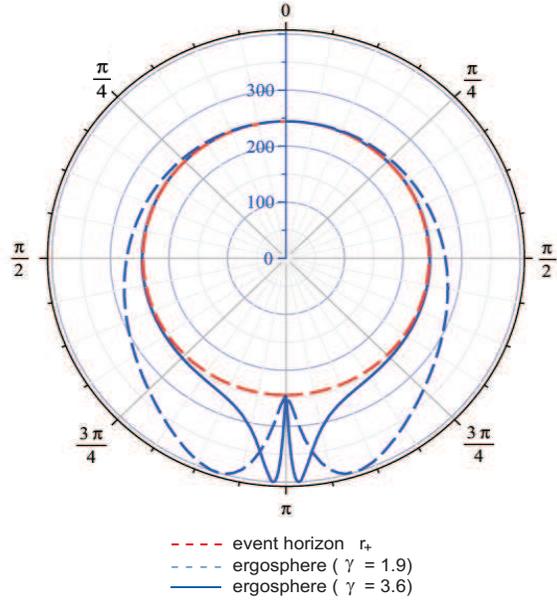}
\caption{The structure of the boosted Kerr black hole for mass $m=200$
and rotation parameter $\omega=195$. The presence of the boost turns the ergosphere
north-south asymmetric as shown by the dominant lobes for $\gamma=1.9$
(dashed blue curve) and $\gamma=3.6$ (continuous blue curve).
The event horizon $r_{+}$ is not altered by the boost (dashed red curve).
The direction of the boost is along the positive $z$-axis ($\gamma > 0$).}
\label{fig:1}
\end{center}
\end{figure}

 The region between the surfaces $r_{\rm stat}(\theta)$ and $r_{+}$ is the ergosphere, where
the Penrose process\cite{penrose,mtw} takes place. The ergosphere is deformed by the boost as a
consequence of the corresponding deformation of $r_{\rm stat}(\theta)$, as illustrated in Figure \ref{fig:1},
for a boosted Kerr black hole with mass $m=200$
and rotation parameter $\omega=195$, in geometrical units. The presence of the boost makes
the ergosphere north-south asymmetric with dominant lobes in the direction opposite to the
boost ($v=\tanh \gamma$).  The deformation increases as the boost increases.
However we can see that the event horizon $r_{+}$ is not altered by the boost.
The direction of the boost is along the positive $z$-axis ($\gamma > 0$).
\section{The singularity}
\label{sec:2}
From the Kretschmann curvature invariant of the spacetime we
have that the metric (\ref{eqn23}) is truly singular at
\begin{eqnarray}
\label{eqn26}
r^2+\Sigma^{2}(\theta)=0.
\end{eqnarray}
The nature of the singularity can be analyzed by transforming (\ref{eqn23}) into the Kerr-Schild form
via the transformation
\begin{eqnarray}
\label{eqn27}
x+iy=(r-i \omega)e^{i \phi} \sin\theta/K(\theta),~~~~z=r~ (\frac{b+a \cos \theta}{a+b \cos \theta}),~~~~u=t-r \;,
\end{eqnarray}
where $r$ is defined by
\begin{eqnarray}
\label{eqn28}
r^4-(R^2-\omega^2) r^2-\omega^2 z^2=0,~~~~R^2=x^2+y^2+z^2 \;,
\end{eqnarray}
so that in this coordinate system the singularity $(r=0,\cos \theta =-b/a)$ corresponds to the
circle
\begin{eqnarray}
\label{eqn29}
x^2+y^2=\omega^2 a^2 \sin^2(\theta_S),~~\theta_S=\arccos (-b/a)
\end{eqnarray}
located at the plane $z=0$. A careful manipulation of (\ref{eqn29}) results in $x^2+y^2=\omega^2$, showing that
the singularity is not altered by the boost, analogously to the event and Cauchy horizons. In this way
the core of the black hole structure {\bf --} the event horizon, the Cauchy horizon and the singularity {\bf --}
does not alter, being independent of the boost parameter. This establishes the invariance of
the black hole structure under Lorentz boosts of the Bondi-Metzner-Sachs group at
future null infinity\cite{sachs1,bondi1}.

\section{Final comments and conclusions}
\label{sec:3}

The metric (\ref{eqn23}) derived here, solution of Einstein's vacuum equations, describes the most general configuration
that an astrophysical black hole must have. The solution has three parameters, mass $m$,
rotation $\omega$ and boost $\gamma$,
which are necessary for the description of black holes in nature. In fact
the processes involving the formation of an astrophysical black hole, like the
merger of binary inspirals of unequal mass rotating black holes, imply that
the resulting remnant black hole must be a boosted Kerr black hole with respect to an
asymptotic Lorentz frame at null infinity. This is the case of the recent detection, by the
LIGO Scientific Collaboration\cite{ligo1}, of the gravitational waves emitted by a binary black hole
merger\cite{ligo} with a mass ratio of the order of $0.8$ -- indicating that the resulting remnant black hole
of this collision must be a Kerr black hole boosted along a particular direction, with respect to the
asymptotic Lorentz frame at null infinity where such emission has been detected.
This must also be the case of astrophysical black holes in which a boost can be inherited
by the collapse of astrophysical objects with large bumps and other deformities.
\par The presence of this further parameter does not violate previous
theorems on the uniqueness of the Kerr black hole solution. The core of the black
hole structure of the boosted Kerr solution, namely, the event horizon,
the Cauchy horizon and the ring singularity remain invariant under the introduction
of a boost in the solution.
\par
On the other hand the static limit surface (which defines the ergosphere) and the
ergosphere itself are affected by the boost. The presence of the
boost turns the ergosphere asymmetric, with dominant lobes in the direction
opposite to the direction of the boost. This asymmetry increases as the boost
parameter increases, as shown in Figure \ref{fig:1}. This asymmetry can
have possible consequences for astrophysical mechanisms connected with the
Penrose processes in the asymmetric ergosphere. Among these processes are the
magnetohydrodynamic Blandford-Znajek mechanisms\cite{blandford,koide} by
which rotational energy can be extracted of Kerr black holes
in the form of electromagnetic and kinetic energy. We argue that
the application of such mechanisms as engines of relativistic jets from
quasars, pulsars and AGNs should be properly treated and
modified in the neighborhood of a boosted Kerr black hole (that describes a
general astrophysical black hole) taking into account the
large north-south asymmetry of the ergosphere.

The author acknowledges the partial financial support of CNPq/MCTI-Brazil, through a Research Grant No. 304064/2013-0.

\end{document}